\newlength{\extralineskip}
\documentclass[12pt]{article}

\usepackage{amssymb}

\begin{document}
\begin{titlepage}
\begin{flushright}
          \begin{minipage}[t]{12em}
          \large UAB--FT--557\\
          \large PRL-THEPH-04-34\\
                 January 2004
          \end{minipage}
\end{flushright}
\vspace{\fill}

\vspace{\fill}

\begin{center}
\baselineskip=2.5em
{\large \bf Neutrino magnetic moments and photo-disintegration
of deuterium}
\end{center}

\vspace{\fill}

\begin{center}
{\bf  $^a$J. A. Grifols, $^a$Eduard Mass{\'o} and $^b$Subhendra Mohanty}
\\
\vspace{0.4cm}
     {\em
$^a$ IFAE, Univ. Aut{\`o}noma Barcelona, Barcelona, Spain\\
$^b$Physical Research Laboratory, Ahmedabad 380009, India.}
\end{center}
\vspace{\fill}

\begin{center}
\large Abstract
\end{center}
\begin{center}
\begin{minipage}[t]{36em}
Neutrinos with non-zero magnetic moments can dissociate deuterium
nuclei by a photon exchange, in addition to the weak neutral
current process. We calculate the neutrino-magnetic moment induced
photo-dissociation cross section of deuterium using the equivalent
photon method. This process would contribute extra events to the
neutral current reaction which is observed with
high precision in the salt-phase of SNO experiment. Using the SNO
data and the recent laboratory measurements of the $^7 Be (p,
\gamma) ^8 B$ reaction which give a more precise value of the
solar $^8 B$ flux we find that the neutrino effective magnetic
moment is $\mu_{eff}^2 = (-2.76 \pm 1.46) \times 10^{-16}
\mu_B^2$ which can be interpreted as an upper bound $|\mu_{eff}| <
3.71 \times 10^{-9} \mu_B$ (at $95\% CL$) on the neutrino magnetic
moments.
\end{minipage}
\end{center}

\vspace{\fill}

\end{titlepage}

\clearpage

\addtolength{\baselineskip}{\extralineskip}

  Enormous progress in neutrino
physics has resulted in recent years from the experiments in
Super-Kamiokande, SNO, KamLAND, and other experimental sites. We
now know beyond any reasonable doubt, for instance, that neutrinos
oscillate which implies a non-vanishing neutrino mass spectrum
\cite{bilenky}. The wealth of new experimental input should
provide also new ways of probing and nailing down neutrino
properties other than masses. Electromagnetic static properties
and, in particular, transition magnetic moments are obvious
quantities to be subject to close scrutiny (precisely because
non-zero neutrino masses naturally allow for helicity flip
transitions) \cite{duncan}. In this paper we use the SNO data
\cite{sno} and
the recent laboratory measurements of the $^7 Be (p, \gamma) ^8 B$
reaction \cite{newboron}
to put restrictions on neutrino magnetic moments.

The theoretical calculation of the deuteron break-up cross section
induced by electromagnetic static quantities of the neutrino has
been carried out in \cite{akhmedov}. However, we follow here
another approach which is simpler and is specially suited for the
magnetic moment case (left-right transition amplitudes do not
interfere with Z-exchange). Indeed, we shall use the equivalent
photon approximation. Of course, although the method gives only
approximate results, they are entirely satisfactory for our
purposes, as we have explicitly checked by comparing to the
calculation in \cite{akhmedov}.

The photon-exchange amplitude of the reaction $\nu_i (k) + d(p)
\rightarrow \nu_j(k^\prime) + n (p_n^\prime)+ p ( p_p^\prime) $
can be written as
\begin{equation}
{\cal{M}} = \frac{l^\mu J_\mu}{ q^2}
\end{equation}
where $J^\mu$ is the hadronic current, $q=k-k^\prime$ the momentum
of the exchanged photon and $l^\mu$ the neutrino current given
explicitly by
\begin{equation}
l^\mu =  \mu_{ij}\,\frac{e}{ 2 m_e} \bar u_i(k^\prime) \sigma^{\mu
\nu} u_j(k) q_\nu
\end{equation}
where $\mu_{ij}$ is the transition magnetic moment
(in units of Bohr-magneton $\mu_{B}= e/(2 m_e)$) of the neutrino mass
eigenstates
involved in the scattering process.
The differential cross section can be written as
\begin{eqnarray}
d\sigma_{mag} = \frac{\mu_{eff}^2}{4 I} \frac{e^2}{4 m_e^2}
\frac{1}{(q^2)^2} \frac{d^3 k^\prime}{(2 \pi)^3 2 k^\prime_0}
 \int ( {l^\mu}^\dagger l^\nu {J_\mu}^\dagger
J_\nu)_{avg}\,  (2 \pi)^4 \delta^4 ( k+p
-k^\prime -p^\prime_n-p^\prime_p)\, d\Pi'
\label{dsigma1}
\end{eqnarray}
\begin{equation}
 d\Pi' \equiv   \frac{d^3 p^\prime_n}{(2 \pi)^3 2
{p^\prime_{0 n}} }\frac{d^3 p^\prime_p}{(2 \pi)^3 2 {p^\prime_{0
p}} }
\end{equation}
where $\mu_{eff}^2$ will be defined later, see
 (\ref{mu_eff_msw}). In (\ref{dsigma1}),
$I \simeq k \cdot p$ is the incident flux and we  have neglected
the neutrino mass in the kinematics. The subscript {\it avg} in
(\ref{dsigma1}) refers to spin-average. The spin-averaged neutrino
current tensor can be explicitly evaluated and turns out to be
\begin{equation}
N^{\mu \nu} \equiv ({l^\mu}^\dagger l^\nu)_{avg} = q^2 ( q^\mu
q^\nu + 2 ( {k^\prime}^\mu k^\nu +{k^\prime}^\nu k^\mu))
\end{equation}
The hadronic current tensor can be written in the general form
\begin{eqnarray}
D_{\mu \nu} \equiv (J_\mu^\dagger J_\nu)_{avg} = a\
\Big( \frac
{-q^2}{p\cdot q} p_\mu p_\nu- p\cdot q\, g_{\mu \nu}
+ p_\mu q_\nu + p_\nu q_\mu \Big)+
b\ ( q^2 g_{\mu \nu} - q_\mu
q_\nu)
\end{eqnarray}
where we have used current conservation
\begin{equation}
q^\mu D_{\mu \nu}  =  q^\nu D_{\mu \nu} =0
\end{equation}
and where $a$ and $b$ are in general functions of the invariants
$q^2, p^2$ and $p\cdot q$. Contracting the spin averaged currents
$N^{\mu \nu}$ and $D_{\mu \nu}$ we get
\begin{eqnarray}
N^{\mu \nu}D_{\mu \nu} = 4 (q^2)^2 \frac{(p \cdot k)^2}{p \cdot
q}\Big[a\ (-1 + \frac{p\cdot q}{p \cdot k})
%\nonumber \\
- b\ \frac{1}{4} \frac{p
\cdot q}{(p \cdot k)^2} q^2 \Big]
 \label{nd}
\end{eqnarray}
 Using the kinematic relations $p=(M_d,\vec 0)$, $k=(E_\nu, \vec k)$,
 $k^\prime=(E_\nu^\prime, \vec k^\prime)$, $q^2= - 2 E_\nu E_\nu^\prime
 (1-cos \theta_{\nu \nu^\prime})$ we find that the coefficient of the $b$ term
 in (\ref{nd}) is
 \begin{equation}
\frac {(E_\nu-E_\nu^\prime)}{M_d}\sin^2
{\theta_{\nu\nu^\prime}\over 2}
\end{equation}
times the coefficient of the $a$ term.
 Since we are dealing with neutrinos in the energy range of $(5-
 20)$ MeV which is much smaller than the deuterium mass, we can
 drop the $b$ term in (\ref{nd}). Substituting in (\ref{dsigma1}) we find that
\begin{eqnarray}
d\sigma_{mag} = \mu_{eff}^2 \frac{e^2}{4 m_e^2}
 \frac{d^3 k^\prime}{(2 \pi)^3 2 k^\prime_0}
 \int \frac{p \cdot k}{p \cdot q}\, a\ (-1 +
\frac{p\cdot q}{p \cdot k})  \,
(2 \pi)^4
\delta^4 ( k+p -k^\prime -p^\prime_n-p^\prime_p)\,  d\Pi'
\label{dsigma2}
\end{eqnarray}
The amplitude for the photo-dissociation process $\gamma(q) + d(p)
\rightarrow n(p^\prime_n) + p(p^\prime_p) $ is
\begin{equation}
{\cal{M}} = \epsilon^\mu J_\mu( q^2=0)
\end{equation}
and the deuterium photo-dissociation cross section can be written
as
\begin{eqnarray}
\sigma_{\gamma} = \frac{1}{4 p \cdot q}
 \int\frac{-1}{2} g^{\mu \nu} ( {J_\mu}^\dagger(0)
J_\nu(0))_{avg} \,
(2 \pi)^4 \delta^4 (
q+p -p^\prime_n-p^\prime_p) \,  d\Pi'
\end{eqnarray}
where again {\it avg} stands for spin average.
Using the fact that $g^{\mu \nu} D_{\mu \nu}(q^2=0) = - 2 a~ p
\cdot q$ we have
\begin{eqnarray}
\sigma_{\gamma} = \frac{1}{4}
 \int a(q^2=0) ~(2 \pi)^4 \delta^4 ( q+p -p^\prime_n-p^\prime_p)
 \,  d\Pi'
\label{sigma4}
\end{eqnarray}
Comparing the neutrino cross section (\ref{dsigma2}) with the
photo-dissociation cross section (\ref{sigma4}) we see that we
have the relation
\begin{eqnarray}
\sigma_{mag} = \frac{ e^2 \mu_{eff}^2}{m_e^2}\int \frac{d^3
k^\prime}{(2 \pi)^3 2 E_\nu^\prime} \Big( \frac{k \cdot p}{q \cdot p}
-1 \Big) \,
\sigma_{\gamma}(q_0=E_\nu -E_\nu^\prime)
\label{sigma5}
\end{eqnarray}
if we assume that in the hadronic current $a\simeq a\mid_{q^2=0}$.
This is totally justified since the transferred momenta are much
smaller than the typical hadronic energies that set the scale
about and beyond which form factors cease to have a point-like
behavior \cite{amalio}. Using the relations $k^\prime_0=|\vec
k^\prime|=E_\nu^\prime$ and using the fact that
\begin{equation}
\frac{k \cdot p}{p \cdot p}-1=\frac{E_\nu^\prime}{E_\nu
-E_\nu^\prime}
\end{equation}
is independent of $\theta_{\nu \nu^\prime}$ we can reduce the
integral $d^3 k^\prime/k_0^\prime = 4 \pi E_\nu^\prime
~dE_\nu^\prime$. The limits of the integration of the variable
$E_\nu^\prime$ are $(0, E_\nu-\epsilon_b)$ where $
\epsilon_b=2.224 $ MeV is the binding energy of deuterium.
Defining the dimensionless variable $x \equiv
(E_\nu-E_\nu^\prime)/E_\nu$ the expression (\ref{sigma5}) for the
neutrino magnetic moment induced deuterium disintegration reduces
to the form
\begin{eqnarray}
\sigma_{mag} = \mu_{eff}^2 \, \frac{\alpha}{\pi}
\left(\frac{E_\nu}{m_e}\right)^2 \int_{\epsilon_b/E_{\nu}}^1 dx
\frac{(1-x)^2}{x}
\, \sigma_\gamma( E_\gamma= x E_\nu)
\label{sigmamag}
\end{eqnarray}
For the photo-dissociation cross section we use the expression
\footnote{Adding the small M1 component of the cross section does
not modify our results appreciably.}
\begin{equation}
\sigma_\gamma(E1)= \sigma_0 \left[ \frac{\epsilon_b(E_\gamma
-\epsilon_b)}{E_\gamma^2} \right]^{3/2}
\label{sigmagamma}
\end{equation}
where the energy dependent factor shown in the square bracket is
the theoretical prediction appropriate for the electric dipole
transition of deuterium \cite{segre}. We have determined the
pre-factor $\sigma_0=19.4$ mb by doing a least square fit of the
energy dependent function shown in (\ref{sigmagamma}) with the
experimental results \cite{d-expt} for the photo-disintegration of
deuterium in the  energy range $E_\gamma \simeq (5-10)$ MeV
appropriate for the $^8 B$ neutrinos observed at SNO.

 The neutrino flux from $^8 B$ in the Sun which is observed
at SNO can be represented by
\begin{equation}
\phi_B(E_\nu) = \Phi_{SSM} ~\xi(E_\nu)
\end{equation}
where $\Phi_{SSM}=(5.87\pm 0.44) \times 10^6$ cm$^{-2}$ sec$^{-1}$
is the new predicted value of the $^8B$ neutrino flux in the
standard solar model \cite{bahcall} after taking into account the
recent laboratory measurements of the $^7Be (p \gamma) ^8B$ cross
section \cite{newboron,gai}. The spectral shape of the $^8 B$ neutrino
flux can be parameterized by the analytical expression
\cite{raffelt}
\begin{equation}
\xi(E_\nu)= 8.52\times 10^{-6}(15.1 -E_\nu)^{2.75} E_\nu^2
\end{equation}
where the neutrino energy $E_\nu$ is in units of MeV. The total
events of deuterium dissociation observed at SNO is the sum of the
standard neutral current events plus those due to neutrino
magnetic moments \footnote{The sum is incoherent because the
magnetic transition amplitude and the weak amplitude do not
interfere.}
\begin{eqnarray}
N^{exp} = N_d  T \Phi_{SSM} \int dE_\nu~ \xi(E_\nu)
%\nonumber\\ &\times&
\, (\sigma_{NC}(E_\nu) + \sigma_{mag}(E_\nu)) \label{nexp}
\end{eqnarray}
where $N_d$ is the total number of deuterium atoms in the fiducial
volume and $T$ is the exposure time. The neutral current flux
reported by SNO \cite{sno} assumes that the total dissociation
events arise from the standard neutral currents,
\begin{equation}
\Phi_{NC}^{SNO} \equiv \frac{N^{exp}}{ N_d T  \int dE_\nu~
\xi(E_\nu) \sigma_{NC}(E_\nu)} ~ \label{phinc}
\end{equation}
Combining (\ref{nexp}) and (\ref{phinc}) we see that the
experimental "NC" flux reported by SNO can be related to the
magnetic moment cross section as
\begin{equation}
\Phi_{NC}^{SNO}= \Phi_{SSM} \left( 1+ \frac{\int dE_\nu ~\xi~
\sigma_{mag} }{\int dE_\nu~  \xi ~\sigma_{NC}} \right)
\label{rate1}
\end{equation}
Factoring out the unknown $\mu_{eff}^2$ from (\ref{sigmamag}), we
can evaluate the numerical factor in the second term of
(\ref{rate1}) by evaluating the integrals over $E_\nu$ in the
range $ (5.5 -15.1)$ MeV. We should add, returning to the comment
we made at the beginning of this paper, that in this energy range
our cross section $\sigma_{mag}$ is smaller than the corresponding
cross section in \cite{akhmedov}. In the higher part of that
energy range, where $\sigma_{mag}$ dominates the integral in
(\ref{rate1}), the difference is about a factor of 2 and thus the
bound on $\mu_{eff}$ we shall obtain below may be considered a
conservative limit by roughly a factor of $\sqrt{2}$. We use the
numerical tables of $\sigma_{NC}(E_\nu)$ given by Nakamura {\it et
al.} \cite{nakamura} to evaluate the denominator of the second
term of (\ref{sigmamag}). We find that the relation between the
experimentally observed flux $\Phi_{NC}^{SNO}$ and the SSM
prediction $\Phi_{SSM}$ is given by
\begin{equation}
\Phi_{NC}^{SNO}= \Phi_{SSM} \left( 1+ 6.06 \times 10^{14}\,
{\mu_{eff}}^2 \right)
\label{rate2}
\end{equation}
The experimental value for the
total neutrino flux assuming the spectral shape of the $^8 B$
neutrinos from the Sun from the recent SNO observations \cite{sno}
is
\begin{equation}
\Phi_{NC}^{SNO} = (4.90 \pm 0.37)\times 10^{6}
{\rm cm}^{-2} {\rm sec}^{-1}
\end{equation}
The fact that the central value of the observed flux is smaller
than the new SSM prediction $\Phi_{SSM} = (5.87 \pm 0.44)\times
10^{6}$ cm$^{-2}$ sec$^{-1}$  leaves room open for the
possibility of sterile neutrinos \cite{smirnov} but it tightens
the constraint on neutrino magnetic moments. Using the numbers
quoted above we find from equation (\ref{rate2}) that
${\mu_{eff}}^2$ is numerically
\begin{equation}
{\mu_{eff}}^2 = (-2.76 \pm 1.46) \times 10^{-16}
\label{musquare}
\end{equation}
where we have added the errors in $\Phi_{SSM}$ and
$\Phi^{SNO}_{NC}$ in quadrature.
 This can be interpreted as an upper bound on $|\mu_{eff}|$ at
$95\%$ C.L. $( 1.96 \sigma)$ given by
\begin{equation}
|\mu_{eff}| < 3.71 \times 10^{-9} \mu_B ~~~(95\% C.L.)
\end{equation}
Earlier bounds on $\mu_{eff}$ \cite{vogel,joshipura} were based on
the extra electron scattering events that can be accomodated by
the SuperK spectrum ($|\mu_{eff}| < 1.5\times 10^{-10}$ at $90\%$ C.L. )
\cite{vogel} and by a combination of all
experimental rates ($|\mu_{eff}| < 2.0 \times 10^{-9}$ at $90\%$ C.L. )
\cite{joshipura}.
Notice that in the case of elastic scattering of electrons, since
the cross section of $\nu_e e^-$
scattering is different from that of $\nu_{\mu,\tau} e^-$, the extra
events due to magnetic moment
scattering were adjusted by the uncertainties in $\delta
 m^2$ and (mainly) $\sin^2 \theta_{12}$.
In our case, since in the deuterium dissociation
neutral current process the cross sections for all three neutrino flavours
are identical, the event rate is
independent of the oscillation parameters and therefore the extra
events due to possible neutrino magnetic moments cannot be accommodated by
shifting the values of the mass squared difference and the mixing angle.
The only extra
parameter with which the magnetic moment can be adjusted is the
theoretical uncertainity in the total $^8 B$ flux.

For the case of the MSW solution of the solar neutrino problem which has
been selected by Kamland \cite{kamland}, the $^8 B$ neutrinos
undergo resonant adiabatic conversion. The matter mixing angle in
the Sun is $\theta_m= \pi/2$. The neutrino mass eigenstate at
production is $\nu_{e} = \nu_{2}$. As the evolution is adiabatic,
at the Earth the neutrinos are still in the $\nu_2$ mass
eigenstate \cite{vogel}. The effective magnetic moment
for the solar $^8 B$ neutrinos is therefore
\begin{equation}
{\mu_{eff} }^2 = {\mu_{21}}^2 + {\mu_{22}}^2 +{\mu_{23}}^2
\label{mu_eff_msw}
\end{equation}
 Our bound on the components of the
neutrino magnetic moment tensor can thus be written as
\begin{equation}
({\mu_{21}}^2 + {\mu_{22}}^2 +{\mu_{23}}^2)^{1/2} < 3.71 \times
10^{-9} \mu_B ~~~(95\%\ {\rm  C.L.})
\end{equation}

At this point a qualification is in order. In fact, in the present
state of affairs one cannot exclude a small contamination of
$\nu_1$ in the neutrinos arriving from the Sun that would depend
on the neutrino energy (see for example \cite{holanda}). We should
emphasize that our bound on $\mu_{eff}$ would still be valid, but
a different interpretation than (\ref{mu_eff_msw}) would follow.
In the future, with more data at hand, it may be worth to
reconsider the interpretation of $\mu_{eff}$ for solar neutrinos.

Sure enough, our bound here is not much different from other
laboratory limits \cite{lab} obtained elsewhere and in fact it is
definitely worse than the one obtained from the plasma emission
argument in globular cluster stars \cite{raffelt}. However, two
facts have to be considered when ascribing it its actual
relevance. First, as we just mentioned the best limit is derived
from energy-loss constraints in stars and hence does rely
exclusively on stellar evolution theory. Second, since neutrinos
oscillate and as a consequence different flavors mix differently
in different settings, reactor, accelerator, solar, and
astrophysical data cannot be compared directly when obtaining the
bounds on magnetic moments \cite{vogel,joshipura}. It is the
analysis of the various pieces of information coming from a
variety of experimental sources that will eventually lead to a
separate restriction on each and every $\mu_{i j}$. The SNO data
used in this paper, and the better data which will hopefully
follow in the future on neutrino initiated deuteron break-up, is
just one source of information among other sources that one can
use to reach this goal.

Work partially supported by the CICYT Research Project
FPA2002-00648, by the EU network on Supersymmetry and the Early
Universe HPRN-CT-2000-00152, and by the DURSI Research Project
2001SGR00188. SM thanks IFAE for  hospitality during the completion of
this work.


\begin{thebibliography}{99}
\bibitem{gai}
We use the number quoted in the last reference in \cite{newboron} for
$\Phi_{SSM}$.
\end{thebibliography}
\end{document}